\documentclass[floatfix,showpacs,amsmath,amsfonts,amssymb,aps,twocolumn, prb,groupedaddress]{revtex4-1}

\usepackage[table,x11names]{xcolor}
\usepackage{graphicx}
\usepackage{dcolumn}
\usepackage{amsthm}
\usepackage{bm}
\usepackage{siunitx}
\usepackage{nicefrac}
\usepackage{todonotes}
\usepackage{mathtools}
\usepackage{float}
\usepackage{braket}
\usepackage{makecell} 
\usepackage{multirow}
\usepackage{wasysym}
\usepackage{titletoc}
\usepackage{gensymb}
\usepackage{multirow}

\definecolor{darkred}{rgb}{0.6,0.,0.}
\definecolor{darkgreen}{rgb}{0.,0.5,0.}
\definecolor{darkblue}{rgb}{0.,0.,0.6}
\usepackage[ breaklinks, colorlinks=true, linkcolor=darkred, citecolor=darkgreen, urlcolor=darkblue]{hyperref}

\newcommand{\includegraphicsgood}{\includegraphics}

\usepackage[inactive,tightpage]{preview}
\PreviewMacro[*[[!]{\input{}}
\PreviewMacro[*[[!]{\includegraphicsgood}

\begin{document}

\title{Fractional quantum Hall states for moir{\'e} superstructures in the Hofstadter regime}
\author{Bartholomew Andrews}
\author{Alexey Soluyanov}
\affiliation{Department of Physics, University of Zurich, Winterthurerstrasse 190, 8057 Zurich, Switzerland}
\date{\today}

\begin{abstract}

We study the transition of $\nu=1/3$ and $2/5$ fractional quantum Hall states of the honeycomb Hofstadter model as we tune to a two-orbital moir{\'e} superlattice Hamiltonian, motivated by the flat bands of twisted bilayer graphene in a perpendicular magnetic field. In doing so, we address the extent to which these states survive in moir{\'e} systems and analyze the nature of the transition. Through the use of a Peierls substitution, we determine the Landau-level splitting for the moir{\'e} Hamiltonian, and study the structure of the Chern bands for a range of magnetic flux per plaquette. We identify topological flat bands in the spectrum at low energies, with numerically tractable lattice geometries that can support the fractional quantum Hall effect. As we tune the model, we find that the orbital-polarized $\nu=1/3$ and $2/5$ states corresponding to the honeycomb Hofstadter model survive up to $\approx 30\%$ of typical moir{\'e} superlattice parameters, beyond which they transition into an insulating phase. We present evidence for this through density matrix renormalization group calculations on an infinite cylinder, by verifying the charge pumping, spectral flow, entanglement scaling, and conformal field theory edge counting. We conclude that fractional quantum Hall states from the Hofstadter model can persist up to hopping amplitudes of the same order as those typical for moir{\'e} superlattice Hamiltonians, which implies generally that fractional states for moir{\'e} superstructures can be discerned simply by analyzing the dominant terms in their effective Hamiltonians.

\end{abstract}

\maketitle         

Moir{\'e} superlattice materials have, for several decades now, served as a test bed for physicists, owing to their unique electronic band structure and exceptional tunability~\cite{Chen19}. In particular, the limelight is on the prototypical moir{\'e} superstructure: twisted bilayer graphene (TBG), which has already had a proven impact on the field~\cite{Bassett58, Dean13}. Moreover, at so-called magic twist angles of TBG, where flat bands are observed at low energies~\cite{Bistritzer11}, it has been shown that it may be possible to engineer low-temperature superconducting phases~\cite{Cao18_1} and ferromagnetism~\cite{Sharpe19}. Perhaps unsurprisingly, such rapid developments and bold claims have also fueled research on the topology of moir{\'e} superstructures, where flat bands are of particular relevance. In many cases, the motivation is to better understand how the fractional quantum Hall (FQH) effect arises, since moir{\'e} systems offer a configurable way of accessing these often elusive and sought-after states. Notably, experimental evidence has already been presented for lattice-generalized FQH states in graphene twisted on a substrate of hexagonal boron nitride~\cite{Spanton18}. Since then there have also been several works on the analytical~\cite{Ledwith19} and numerical~\cite{Abouelkomsan19} theory of FQH states in moir{\'e} superstructures, as well as interest in moir{\'e} systems in a perpendicular magnetic field~\cite{Lu19} and the construction of effective tight-binding models to expedite many-body numerics~\cite{Koshino18}. In particular, there have been many proposals for effective lattice models for such systems of ``two-orbital Hubbard" type~\cite{Zhu19, Venderbos18}. In a perpendicular magnetic field, the dominant term in these models is often the Hofstadter Hamiltonian. Motivated by this, we investigate the extent to which FQH Hofstadter physics persists by tuning from the honeycomb Hofstadter model to a two-orbital Hubbard-type moir{\'e} model in a perpendicular magnetic field.

By examining a recently proposed moir{\'e} superlattice Hamiltonian~\cite{Yuan18, Koshino18}, inspired by the low-energy physics of magic-angle twisted bilayer graphene, in a perpendicular magnetic field, we start by analyzing the band structure as a function of flux per plaquette. From this spectrum, we examine numerically tractable lattice geometries that yield topological bands with a large gap-to-width ratio, and hence the potential to host fractional states~\cite{Bergholtz13}. Subsequently, we fractionally fill the lowest band with spinless fermions in accordance with the generalized Jain series for the Hofstadter model, focusing on the Laughlin FQH state at $\nu=1/3$ filling~\cite{Laughlin83} and the hierarchy state at $\nu=2/5$. We then add up to nearest-neighbor interactions and solve the many-body problem using infinite density matrix renormalization group (iDMRG) calculations on a thin cylinder geometry~\cite{White92, Cincio13, Grushin15}. We tune from the honeycomb Hofstadter model to the full moir{\'e} model in order to study when and how these states break down. These calculations are particularly intensive, since we are taking into account up to fifth-nearest-neighbor hoppings, and two orbitals per site. Using the results from flux insertion and entanglement entropy, we present evidence of the first two hierarchy states in this effective model. Moreover, as we tune from the underlying honeycomb Hofstadter model to the full effective model, we show that these states survive up to $\sim30\%$ of the typical superlattice hopping parameters, which is remarkably deep into the moir{\'e} regime. Finally, we analyze the nature of this transition and uncover the origin of the fractional states for the given lattice geometries. These findings show that it is possible to realize FQH states in the effective model and, furthermore, that the states of the underlying honeycomb Hofstadter model are significant, with the potential to survive up to the same order of magnitude as typical moir{\'e} hopping amplitudes. This suggests that states from dominant underlying Hamiltonians persist for other effective moir{\'e} models.          

This paper is organized as follows. In Sec.~\ref{sec:model}, we introduce and justify the tight-binding model, and detail the many-body numerics employed. In Sec.~\ref{sec:results}, we describe the configurations used to obtain fractional phases on the lattice and present our key results. Finally, in Sec.~\ref{sec:conclusion}, we discuss explanations for the results and give our outlook for future research directions.    

\section{Model}
\label{sec:model}

In this section we present the effective moir{\'e} tight-binding model inspired by the physics of magic-angle TBG in a perpendicular magnetic field. In Sec.~\ref{subsec:single_ham}, we analyze the properties of the single-particle Hamiltonian, and in Sec.~\ref{subsec:many_ham} we add interaction terms to define the many-body Hamiltonian. 

\subsection{Single-particle Hamiltonian}
\label{subsec:single_ham}

We consider an effective two-orbital Fermi-Hubbard model based on the four flat bands of TBG at the first magic angle ($\theta_\text{m}=1.05\degree$) in a perpendicular magnetic field. We consider a spinless model, unlike in a real TBG system, since our motivation is to study fractional phases in moir{\'e} Hamiltonians at a low computational cost. The tight-binding model is defined on the emergent moir{\'e} honeycomb lattice, with $(p_x,p_y)$ orbitals at each site. The single-particle Hamiltonian is given as:
\begin{equation}
\label{eq:single_ham}
\begin{split}
H_0 &= \sum_{\braket{ij}} \left[ t_1 e^{\mathrm{i}\theta_{ij}} \mathbf{c}_i^\dagger \cdot \mathbf{c}_j + \text{H.c.} \right] \\
&+ \kappa\sum_{\braket{ij}_5} \left[ t_2 e^{\mathrm{i}\theta_{ij}} \mathbf{c}_i^\dagger \cdot \mathbf{c}_j + t_2' e^{\mathrm{i}\theta_{ij}} (\mathbf{c}_{i}^\dagger \times \mathbf{c}_{j})_z + \text{H.c.} \right],
\end{split}
\end{equation}
where $\mathbf{c}^\dagger=(c^\dagger_x, c^\dagger_y)^\intercal$ is the fermion creation operator for the $(p_x,p_y)$ orbitals, $\theta_{ij}$ is the Peierls phase in moving from site $i$ to site $j$, $\braket{}_n$ denotes the $n$th-nearest neighbors, $(t_1, t_2, t_2')=(1, -0.025, 0.1)$~meV are taken as the typical hopping amplitudes~\cite{Yuan18, ErratumYuan18, Koshino18}~\footnote{Specifically, the tight-binding parameter magnitudes are taken from the erratum of the paper by Yuan \& Fu~\cite{ErratumYuan18} and normalized such that $t_1=1$.}, and $\kappa$ is a dimensionless tuning parameter.

The form of the minimal tight-binding model at zero magnetic field was first defined by Yuan and Fu in 2018, based on symmetry arguments, to capture the physics of the metal-insulator transition in magic-angle TBG~\cite{Yuan18}. Later, the scope of the model was made more precise by Koshino~\textit{et~al.}, by showing that an extended superposition of Wannier functions can exactly recover the dispersion of the flat bands~\cite{Koshino18}. We note that Po~\textit{et~al}. have since argued that more bands in the tight-binding model are required to accurately recover the symmetry representations and topology of the magic-angle TBG system~\cite{Po19}, and hence we stress that our base model is only realistic to the extent outlined in the cited literature~\cite{Yuan18, Koshino18}. For the purposes of this paper, we choose this model as a numerically accessible way to showcase the principals and methodology behind realizing FQH states in moir{\'e} superlattice Hamiltonians.

The $t_1$ term describes same-orbital nearest-neighbor (A$\to$B or B$\to$A) hopping on the honeycomb lattice ($\hexagon_1$) and the $t_2$ term describes same-orbital fifth-nearest-neighbor (A$\to$A or B$\to$B) hopping ($\hexagon_5$); or equivalently, the second-nearest-neighbor hopping within the triangular sublattice ($\triangle_2$). Together, these two terms describe the minimal tight-binding model for equal-amplitude $p$-orbital hoppings on a honeycomb lattice\footnote{This is also the reason why intra-sublattice 2\textsuperscript{nd}-nearest neighbor hopping would not suffice.} and are $SU(4)$ symmetric, for the original spin-degenerate eight-band model. Note that the $t_2$ term (weakly) breaks the particle-hole symmetry. The $t_2'$ term ($\hexagon'_5$) represents the orbital mixing. This term lifts the orbital degeneracy along the $\Gamma$$\to$M line in the band dispersion, and it reduces the symmetry of the Hamiltonian down to $U(1)\times SU(2)$, where $U(1)$ is for orbital chirality and $SU(2)$ is for spin (in the spin-degenerate model). The chiral basis is given as $c_{\pm}=(c_x\pm \mathrm{i}c_y)/\sqrt{2}$ for the $p_x\pm\mathrm{i}p_y$ orbitals. By matching to the effective continuum model~\cite{Santos07, Bistritzer11}, the first- and fifth-nearest-neighbor hopping terms were shown to be the dominant contributions for the inter- and intra-sublattice hoppings, respectively~\cite{Koshino18}. Moreover, these terms form the minimal model that satisfies the symmetry constraints and reproduces the key features of the band structure. Further details of the single-particle Hamiltonian at zero magnetic field are discussed in the papers by Yuan and Fu~\cite{Yuan18}, as well as Koshino~\textit{et~al.}~\cite{Koshino18}. We reiterate at this point that the limitations of the model used in this paper are largely due to the motivation to apply many-body numerical calculations. The accuracy of the model in representing moir{\'e} superstructures can naturally be improved by including more terms, with longer-range hoppings, or by directly applying a Peierls substitution to the continuum model, as in Bistritzer and MacDonald~\cite{Bistritzer11_2}. Here we focus on the simplest suggested lattice Hamiltonian to make progress~\cite{Yuan18, Koshino18}. The most important properties of this base model are that it has terms to account for inter-sublattice, intra-sublattice, and orbital-mixing hopping; it recovers the orbital-degeneracy lifting of the energy bands along the $\Gamma$$\to $M line and the trefoil form of the electronic Wannier orbital; and it is the minimal model that can be constructed by such a Wannierization~\cite{Koshino18}.     

\begin{figure}
	\includegraphics[width=\linewidth]{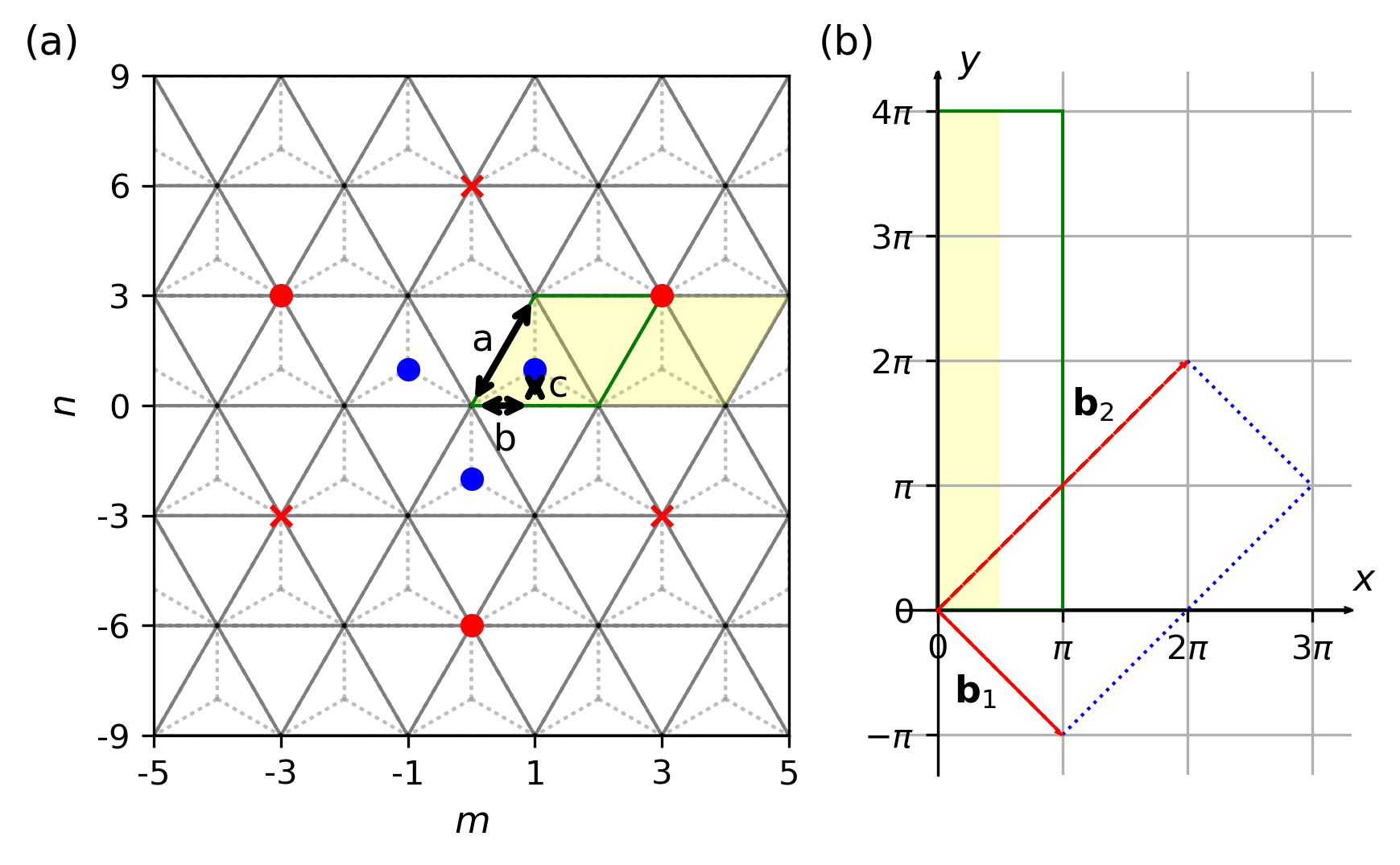}
	\caption{\label{fig:lattice_bz} (a)~Real-space lattice in units of $(x,y)=(mb,nc)$, where $b\equiv a/2$, $c\equiv(1/3)(\sqrt{3}a/2)$, and $a\equiv 1$ is the lattice constant of the triangular sublattice or, equivalently, the length of the hexagonal basis vector. The coordinates of the nearest neighbors ($\hexagon_1$) are shown in blue, and the coordinates of the fifth-nearest neighbors ($\hexagon_5$) are shown in red. The two conjugate sets of three for the fifth-nearest neighbors are distinguished by dots and crosses. The unit cell is outlined in green and an example of a $q=2$ magnetic unit cell is highlighted in yellow. (b)~Sketch of Brillouin zone using the coordinates $x=k_x a/2$ and $y = \sqrt{3}k_y a/2$, with reciprocal-lattice vectors $\mathbf{b}_1$ and $\mathbf{b}_2$. The green outline shows the Brillouin zone (effective reciprocal unit cell) over which the Chern number calculations are performed, and the yellow region shows the $q=2$ magnetic unit cell corresponding to (a).}
\end{figure}

We apply a perpendicular magnetic field to our base system, $\mathbf{B}=B\hat{\mathbf{e}}_z$, via a Peierls substitution~\cite{Peierls33} with Peierls phases given as $\theta_{ij}=(2\pi/\phi_0)\int_i^j \mathbf{A}\cdot\mathrm{d}\mathbf{l}$, where $\phi_0$ is the flux quantum, $\mathbf{A}$ is the vector potential and $\mathrm{d}\mathbf{l}$ is an infinitesimal line element. For the Peierls analysis, we work in the Landau gauge in the $x$ direction, such that $\mathbf{A}=Bx\hat{\mathbf{e}}_y$. Figure~\ref{fig:lattice_bz} shows a sketch of the lattice model in Eq.~(\ref{eq:single_ham}) along with its unit cell. When a magnetic field is applied, the unit cell is extended to relative dimensions of $q \times 1$ in real space, and the Brillouin zone is correspondingly contracted. In this analysis, we define the magnetic flux passing through a unit cell as the flux density $n_\phi \equiv BA_\text{UC}/\phi_0\equiv p/q$, where $A_\text{UC}=\sqrt{3}a^2/2$ is the area of a unit cell (shown in Fig.~\ref{fig:lattice_bz}); $\phi_0=h/e$ is the flux quantum; and $p$ and $q$ are coprime integers. In the Landau gauge, $q$ is the area of the magnetic unit cell, and $p$ corresponds to the splitting of the lowest Landau level~\cite{Hofstadter76}. In the continuum limit ($n_\phi\to 0$), Landau-level physics is recovered, whereas away from this limit Landau-level physics is generalized to lattice-based systems. The details of the Peierls substitution are discussed in Appendix~\ref{sec:peierls}.

\begin{figure}
	\includegraphics[width=\linewidth]{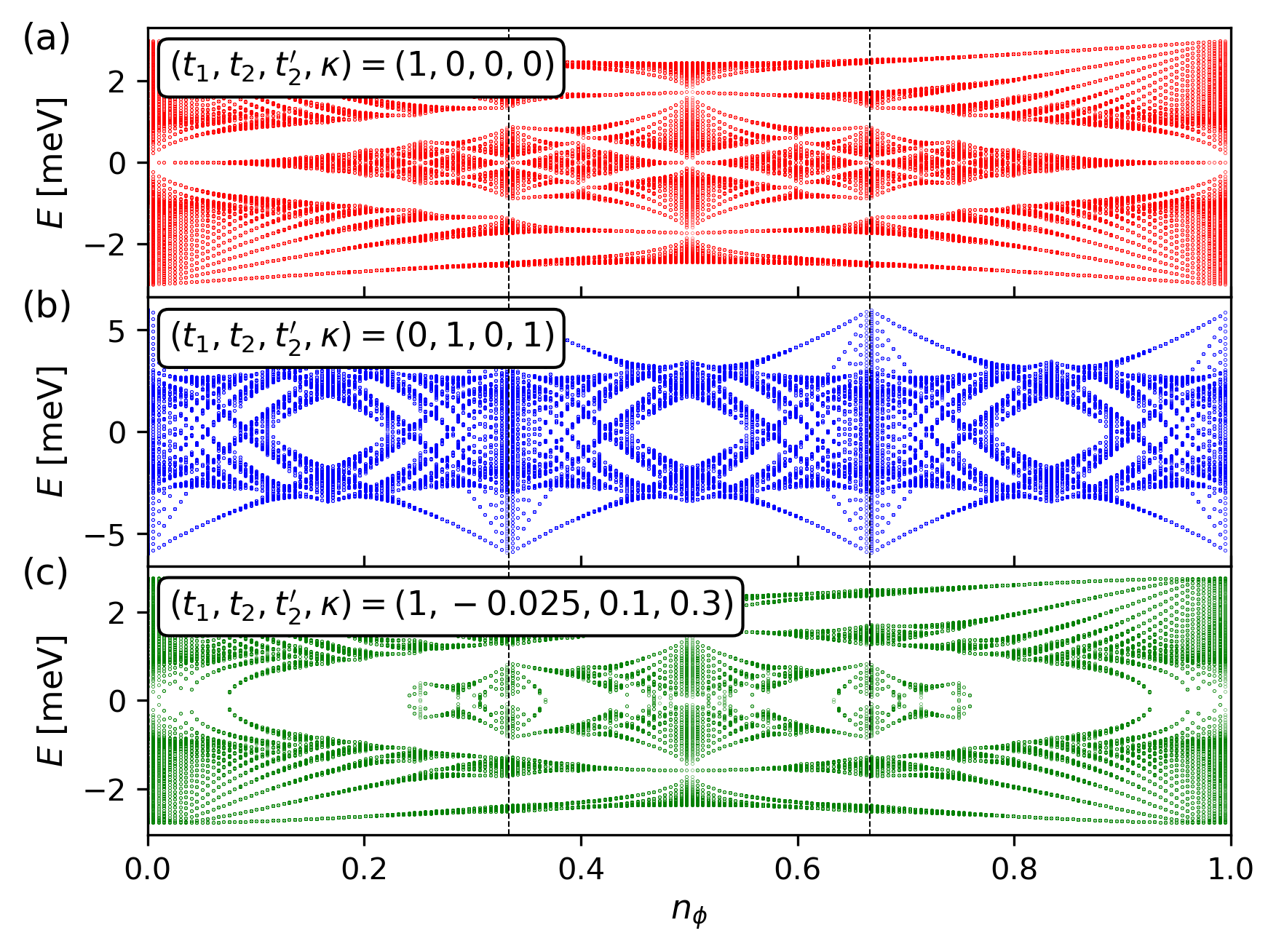}
	\caption{\label{fig:butterfly} Energy spectrum of Eq.~(\ref{eq:single_ham}) as a function of flux density $n_\phi=p/q$. $p$ and $q$ are coprime integers with $q=199$ and $0\leq p \leq 198$. We consider the cases of (a) $\hexagon_1$ hopping with $(t_1, t_2, t_2')=(1, 0, 0)$~meV, (b) $\hexagon_5$ hopping with $(t_1, t_2, t_2')=(0, 1, 0)$~meV, and (c) typical hopping parameters for the model: $(t_1, t_2, t_2')=(1, -0.025, 0.1)$~meV~\cite{Yuan18, ErratumYuan18, Koshino18} with $\kappa=0.3$. Since the bandwidths are narrow at the flux densities considered, we plot the points at $(k_x,k_y)=(0,0)$ only. The dashed lines at $n_\phi=1/3$ and $2/3$ indicate the high-symmetry points in the $\hexagon_5$ Hamiltonian.}
\end{figure}

The single-particle energy spectrum of the model as a function of flux density is presented in Fig.~\ref{fig:butterfly}. As in the Hofstadter model, the (in)commensurability between the two area scales defined by the flux quantum and the magnetic unit cell gives rise to a fractal spectrum with an infinite selection of Chern bands. Figure~\ref{fig:butterfly}(a,b) shows that the model recovers the expected energy spectra for the purely $\hexagon_1$~\cite{Rammal85} and $\hexagon_5$~\cite{Oh00} hoppings, respectively. Figure~\ref{fig:butterfly}(c) shows the energy spectrum for Eq.~(\ref{eq:single_ham}) with typical tight-binding parameters based on the references where the model was introduced~\cite{Yuan18, ErratumYuan18, Koshino18} and also serves as our first original result\footnote{The motivation for this choice of $\kappa=0.3$ is revealed in Sec.~\ref{sec:results}. It is the largest value of $\kappa$ for which the FQH states from the Hofstadter model are observed.}. Some clearly defined Chern sectors from the simpler systems (e.g., $\hexagon_1$~\cite{Agazzi14}) are now overlapping, which obstructs a clear general identification, e.g., via a Diophantine relation. Instead, the Chern numbers of the bands need to be computed on an individual basis. It can also be seen that the symmetric points of the $\hexagon_1$ and $\hexagon_5$ spectra (at multiples of $n_\phi=1/3$) are identifiable for the example case in Fig.~\ref{fig:butterfly}(c).

\subsection{Many-body Hamiltonian}
\label{subsec:many_ham}

We extend our model to the interacting case through the addition of on-site and nearest-neighbor density-density terms, such that the full Hamiltonian becomes
\begin{equation}
\label{eq:many_ham}
H = H_0 + U\sum_i \rho_{x,i} \rho_{y,i} + V \sum_{\braket{ij}} \rho_i \rho_j,
\end{equation}
where $U$ and $V$ are the on-site and nearest-neighbor interaction strengths, $\rho_{x(y)}=c_{x(y)}^\dagger c_{x(y)}$ is the spinless fermion density operator for an $x(y)$ orbital, and $\rho=\mathbf{c}^\dagger\cdot\mathbf{c}$ is the total density operator. Since the electronic Wannier orbitals in our model extend over unit cells, interaction effects play an important role in the many-body physics. Furthermore, due to typical screening by metallic gates in realistic graphene/moir{\'e} devices, nearest-neighbor interactions ($\approx10$ nm) are expected to be sufficient to model this effect~\cite{Yuan18,Koshino18}. Precisely determining the strength of effective Coulomb interactions in lattice models for magic-angle TBG, or generally flat-band moir{\'e} superstructures, is an ongoing area of research~\cite{Tommaso20, Vanhala19, Pizarro19, Zhang20, Rademaker18}. In this paper, we use $U=10V=100 t_1$ as an order-of-magnitude estimate, however we note that this is somewhat large compared to the most recent consensus, which lies in the range of {$10$--$40$~meV~\cite{Tommaso20, Zhang20, Pizarro19}}.  

In order to tune to a fractionalized phase in our system, there are several independent conditions that first need to be satisfied. For FQH states, the interaction effects need to dominate over the kinetic energy. In practice, this is achieved for isolated topological flat bands in the energy spectrum, where the density of states is (near) singular and there is a large energy cost for fermions to jump to the next-highest band(s)~\cite{Parameswaran13}. Second, these flat bands need to be partially filled, with a desired filling fraction compatible to the lattice geometry and flux density. For Abelian FQH states, these filling fractions are defined by the Jain series~\cite{Jain89} (and for inherently lattice-based states by a generalized Jain series~\cite{Moller15}). Finally, the interaction strength for the fermions in the partially filled isolated topological flat band(s) needs to be large compared to the band width, $W$, such that the interactions dominate over the kinetic energy, but small compared to the band gap, $\Delta$, such that the fermions do not hop to the next band: $W\ll V \ll \Delta$~\cite{Bergholtz13}. In practice, it has been shown that FQH states can persist in some cases even when the second inequality is relaxed~\cite{Regnault11, Kourtis14}. The details of example lattice geometries required to realize the FQH effect are presented in Sec.~\ref{subsec:lat_geom}.   

We solve the many-body problem numerically using the iDMRG algorithm on a thin cylinder geometry~\cite{White92, Cincio13, Grushin15}. Initially, the Hamiltonian is transcribed to a matrix product operator (MPO) representation and the wave function is represented as a matrix product state (MPS), which is wound around a cylinder and bipartitioned into left and right parts. The MPS representation yields a Schmidt decomposition ${\ket{\psi}=\sum_{\alpha}\Lambda_{\alpha} \ket{\alpha_\mathrm{L}}\otimes\ket{\alpha_\mathrm{R}}}$ on each bond, up to bond dimension $\chi$, where $\Lambda_{\alpha}$ and $\ket{\alpha_{\mathrm{L}/\mathrm{R}}}$ are the Schmidt coefficients and left/right Schmidt states. These two quantities are directly related to the eigensystem of the reduced density matrix $\rho^{\mathrm{L}/\mathrm{R}}_{\alpha}$, and hence the entanglement spectrum~\cite{Li08}. Specifically, the Schmidt states correspond directly to the eigenstates, and the Schmidt values are the square of the eigenvalues: $\rho_\alpha^{\mathrm{L}/\mathrm{R}}=\Lambda_\alpha^2$. The corresponding (von Neumann) entanglement entropy is defined as $S_\mathrm{vN}=-\sum_\alpha \Lambda_\alpha^2 \log(\Lambda_\alpha^2)$. Iteratively sweeping over bonds in the MPS and performing two-site eigensystem updates until convergence of the energy, entanglement entropy, and other criteria, then yields a matrix product representation of the ground-state wave function.

For the iDMRG algorithm, a thermodynamic limit ansatz is used for the cylinder axis direction and periodic boundary conditions in the azimuthal direction. Overall the procedure is in the semi-thermodynamic limit and can reach system sizes (for the finite circumference) competitive with traditional algorithms, such as exact diagonalization and Monte Carlo. Furthermore, no band projection needs to be taken for the interaction Hamiltonian.

The challenges of implementing an iDMRG calculation for the Hamiltonian in Eq.~(\ref{eq:many_ham}) are considerable, since the hoppings are long range ($\hexagon_5$) and the Hilbert space is large. In this paper, we overcome these issues in a variety of ways. For example, we perform bond dimension convergence analysis to ensure that the low bond dimensions that we are able to achieve are representative for the model and we deploy the code using OpenMP parallelization on a cluster of high-end processors. There are, however, numerous ways in which the calculations could be expedited, which we hope to explore in the future. In particular, we note the massively parallelized implementations of the DMRG algorithm using the Message Passing Interface (MPI) standard~\cite{Kantian19} and adaptations to GPUs~\cite{Milstead19}, which have both demonstrated success in the past year. Further details on the iDMRG simulations used in this paper are found in Appendix~\ref{sec:iDMRG}.

\section{Results}
\label{sec:results}

In this section we present our numerical results, including many-body computations using iDMRG. In Sec.~\ref{subsec:top_flat_bands}, we show the existence of topological flat magnetic sub-bands that are tractable with many-body numerics. In Sec.~\ref{subsec:lat_geom}, we outline the lattice geometries needed for fractional topological phases. Subsequently, in Sec.~\ref{subsec:FCIs} we show evidence for the integer and fractional quantum Hall effect in the flat magnetic sub-bands of this moir{\'e} superlattice Hamiltonian.

\subsection{Topological flat bands}
\label{subsec:top_flat_bands}

\begin{figure}
	\includegraphics[width=\linewidth]{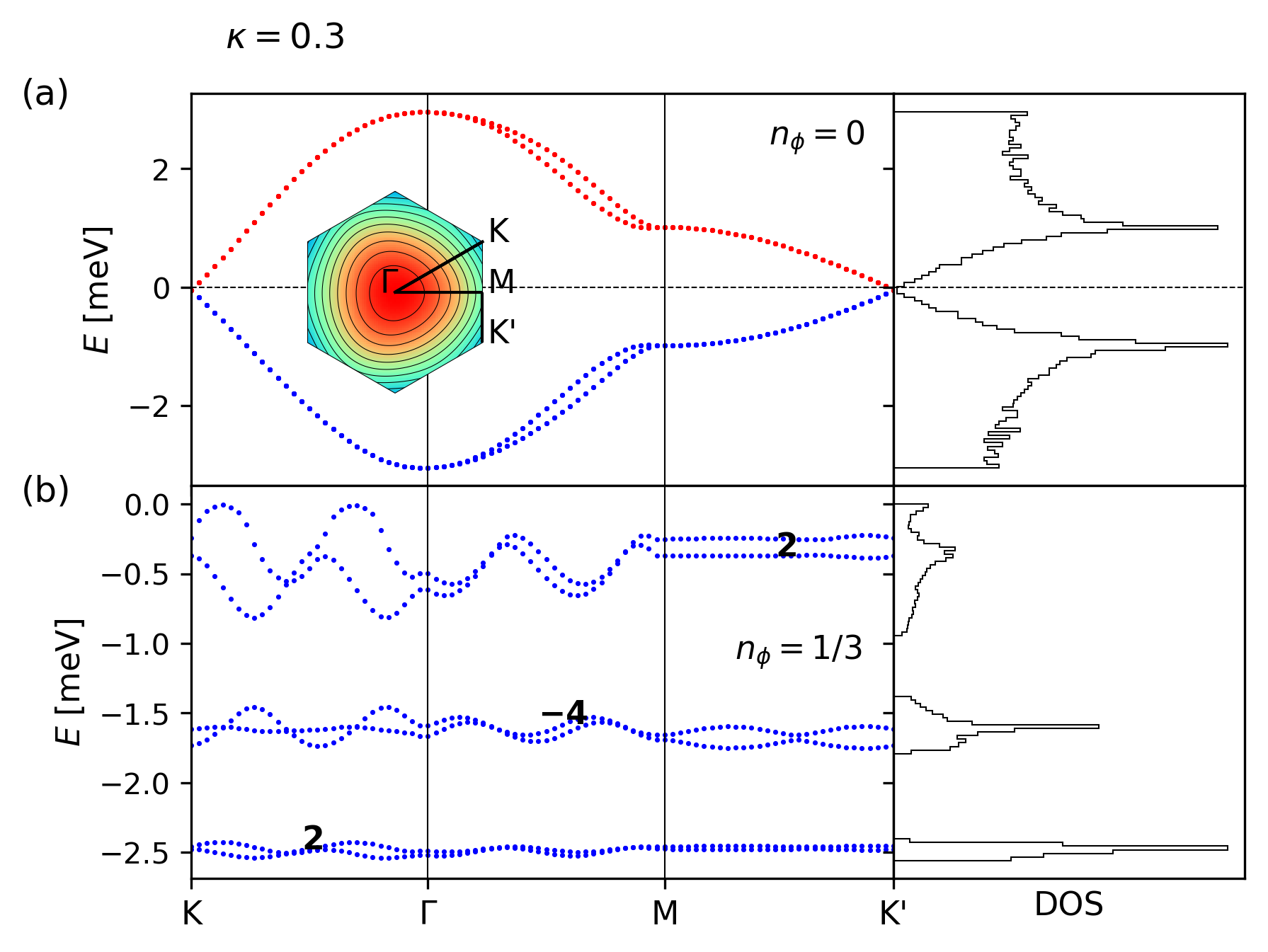}
	\includegraphics[width=\linewidth]{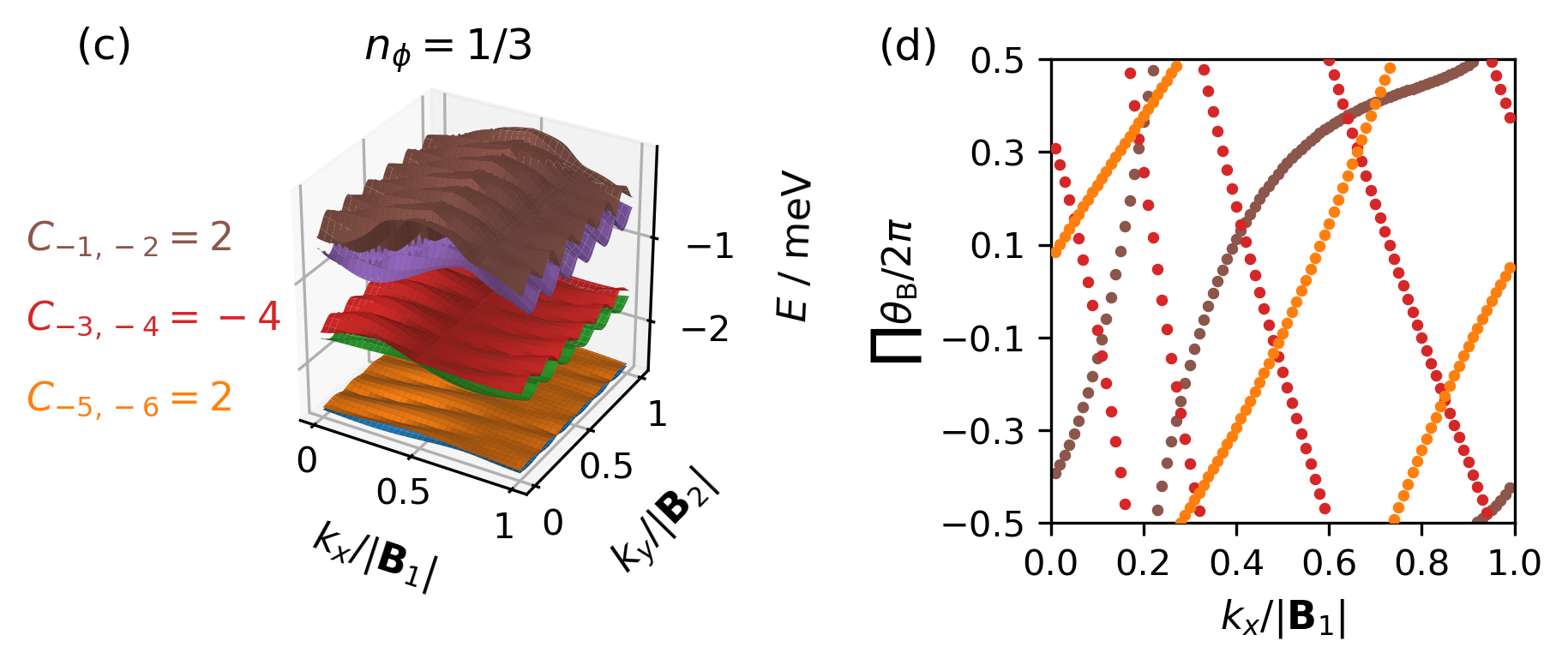}
	\caption{\label{fig:band_structure} Band structure of Eq.~(\ref{eq:single_ham}) at $\kappa=0.3$ with $n_\phi=0$ and $1/3$. (a) Zero-field band structure. The insert shows the Brillouin zone and energy contour plot for the $\xi=+$ valley. (b)~The Landau-level splitting of the lower-half energy bands, labeled by their corresponding Chern numbers. The right panel shows the density of states (DOS), based on a $100^2$ grid with $100$ bins per energy range. (c)~The 3D band structure of the three lowest bands in (b). (d)~The product of Berry phases, $\theta_\mathrm{B}$, around Wilson loops in the $k_y$ direction for the bands shown in (c).}
\end{figure}

At zero magnetic field, the band structure of Eq.~(\ref{eq:single_ham}) takes the form shown in Fig.~\ref{fig:band_structure}(a). Compared to graphene, we note that since the $t_2$ term breaks the particle-hole symmetry the band structure is no longer symmetric about the $E=0$ line. The $\pm$ sign of $t_2$ shifts the spectrum up/down, respectively. Furthermore, since the $t_2'$ term lifts the orbital degeneracy, we notice a band splitting along $\Gamma\to\mathrm{M}$. The inset of Fig.~\ref{fig:band_structure}(a) shows the energy dispersion of the $\xi=+$ valley. From this, we can see a (weakly) trefoil form of the Wannier orbital, which is sharpened as the magnitude of $t_2'$ is increased~\cite{Koshino18}. The electronic Wannier orbital is also seen to span multiple unit cells, which supports the significance of interaction effects in the many-body problem. From an appropriate summation of Bloch states in the effective continuum model, the exact form of the energy dispersion can be calculated~\cite{Koshino18}. The exact dispersion shows an energy maximum at the $\Gamma$ point with narrower lobes of the trefoil dispersion, and a suppressed particle-hole asymmetry. Otherwise, the key features of the energy dispersion hold.  

We now apply a perpendicular magnetic field to our system, with flux density $n_\phi=p/q$. Consequently, our system splits into $4q$ magnetic sub-bands (due to the orbital degree of freedom). The larger the value of $q$, the flatter the bands, at the expense of a larger magnetic unit cell. Therefore, a compromise needs to be reached for the many-body numerics, as well as experimental accessibility. A sequence of flux densities with small $q$ values is given as $n_\phi=1/3,1/4,1/5$ with corresponding gap-to-width ratios of $\Delta/W=3.64, 2.10, 77.6$ at $\kappa=0.3$. The $q=3$ magnetic sub-bands of the lower-half bands in Fig.~\ref{fig:band_structure}(a) are shown in Fig.~\ref{fig:band_structure}(b) for $\kappa=0.3$. As seen from the plot, each pair of bands is isolated and labeled by their corresponding non-interacting Chern number -- the total Chern number of the bands is zero. The lowest pair of bands has a near-singular density of states and a large gap to the next-highest bands, making this flux density a good candidate for hosting fractionalized phases. Figure~\ref{fig:band_structure}(c) shows a three-dimensional plot of the band structure of the lowest three pairs of bands in the reduced Brillouin zone, and Fig~\ref{fig:band_structure}(d) is a polarization plot showing the winding corresponding to the Chern number. From this we can clearly see that the lowest-energy bands are isolated (at all $\mathbf{k}$), topological, and flat.         

\subsection{Lattice geometries}
\label{subsec:lat_geom}

In order to probe the effect of band flatness on the potential topological phases, we systematically examine a variety of lattice geometries. Here, we detail the procedure for selecting a state at a particular filling fraction, and present the numerical results motivating our choice of lattice configurations.

The filling factor of a state is defined as $\nu=n/n_\phi$, where $n$ is the total filling fraction of the system and $n_\phi$ is the flux density, both with respect to lattice unit cells in this paper. The `system' is defined as the dimensions of a magnetic unit cell $(l_x,l_y)$ (in units of lattice unit cells), tiled to fit the total dimensions of $(L_x,L_y)$ (in units of magnetic unit cells). Therefore, the total number of lattice unit cells is given as $N_u=L_x l_x \times L_y l_y$, and the total filling fraction of the system is $n=N/N_u$, where $N$ is the number of particles\footnote{Note that the typical definition of $\nu=N/N_s$ for the square Hofstadter model, where $N_s$ is the number of lattice sites, only holds because $N_s=N_u$. This stems from the fact that we are fractionally filling bands in our spectrum, and the total number of available sites in the unit cell of our Hamiltonian determines the number of bands.}. In this paper, we use the convention that $x$ is along the cylinder axis and $y$ is along the circumference.

\begin{figure}
	\includegraphics[width=\linewidth]{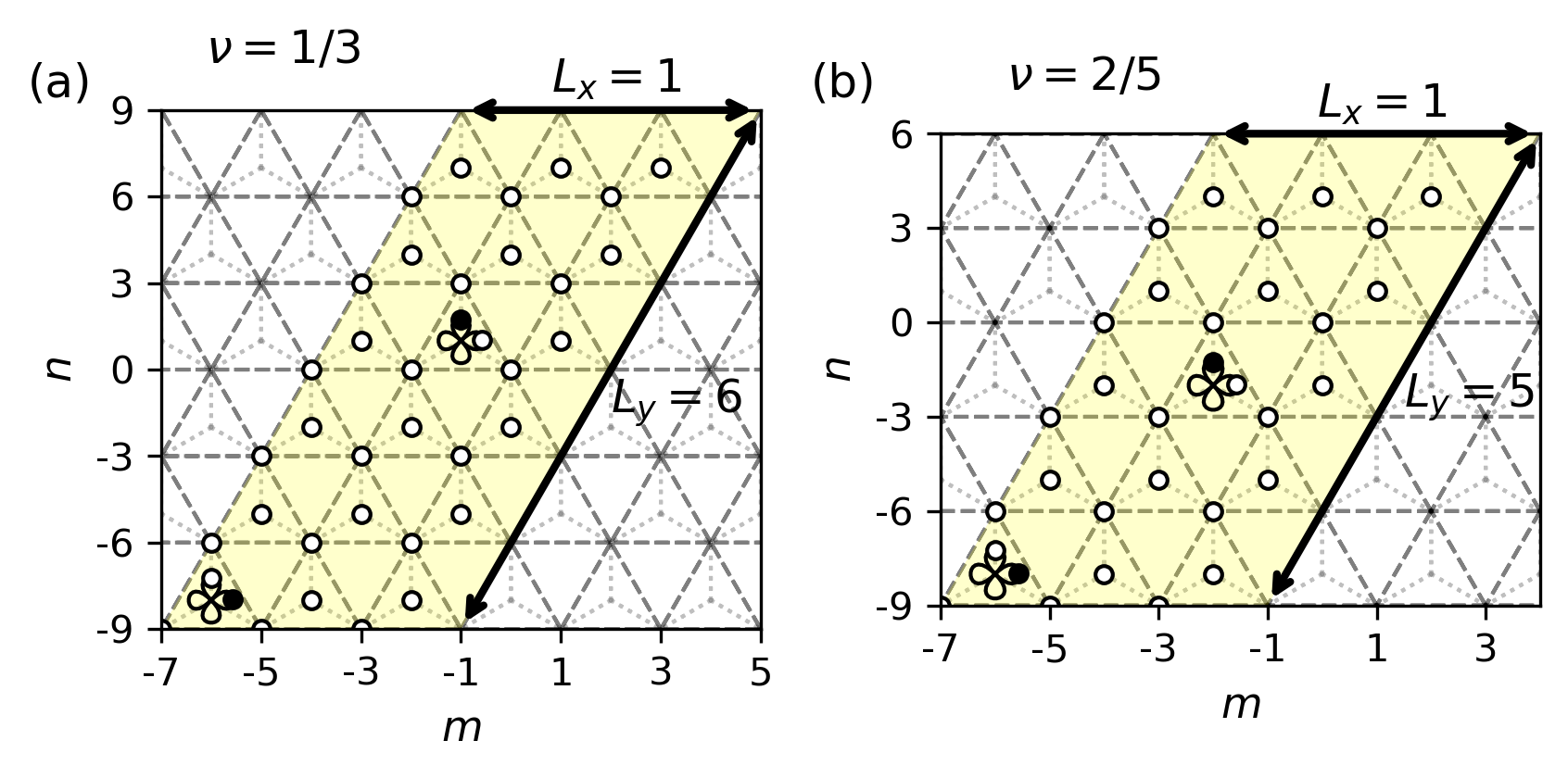}
	\caption{\label{fig:example_geom} Example initial MPS unit-cell geometries for a honeycomb (emergent moir{\'e}) lattice in a perpendicular magnetic field with flux density $n_\phi=1/3$. Empty (full) sites are indicated by white (black) discs and the $p$ orbitals at each site are not drawn unless needed. The MPS filling and geometry are chosen such that the (a) $\nu=1/3$ and (b) $\nu=2/5$ states may be recovered in the moir{\'e} Hamiltonian.}
\end{figure}

The generalized Jain series from composite fermion theory for Abelian FQH states in Chern bands of the Hofstadter model is given as
\begin{equation}
\label{eq:Moller}
\nu=\frac{r}{|kC|r+1},
\end{equation}
where $|r|\in\mathbb{Z}^+$ is the number of fully filled bands in the composite fermion spectrum, $\mathrm{sgn}(r)$ is the sign of the Chern number for the composite fermion band relative to the sign of the Chern number for the low-energy manifold, and $k\in\mathbb{Z}^+$ is the number of flux attached to each vortex (one for bosons and two for fermions)~\cite{Moller15, Andrews18}. The simplest example of a fermionic fractional phase in this hierarchy is therefore the state at $r=1$ and $C=1$, i.e.,~the Laughlin state with $\nu=1/3$~\cite{Laughlin83}. It has been shown experimentally in the case of the Laughlin state for the Hofstadter model that $n_\phi\lesssim0.4$ may be required to obtain stable FQH states\cite{Hafezi07}. In this example, we therefore select $n_\phi=1/3$. Since the honeycomb lattice has two atoms per unit cell, this will yield a magnetic unit cell that is six lattice sites across. We note that a total filling fraction $n=1/9$ of the system with respect to lattice unit cells corresponds to $1/18$ with respect to lattice sites, and $1/36$ with respect to orbital sites. This is due to the fact that the number of lattice sites $N_s=2N_u$ and the number of orbital sites $N_o=2N_s$. We therefore need to construct a system with a total MPS filling of $1/36$ with respect to orbital sites, and at least two particles (for faster convergence). The dimensions of the magnetic unit cell are already fixed by the flux density to be $3\times 1$ and so the remaining freedom is in the system dimensions $(L_x,L_y)$. Since $L_x$ is in the direction of the thermodynamic limit ansatz, it typically suffices to set $L_x=1$\footnote{We note that in some cases a larger $L_x$ may uncover competing states, such as charge density wave states, in the phase diagram. Due to computational constraints, we were not able to investigate this further in this project.}. Consequently, we require $L_y=6$ to host two particles in our system. An illustration of this example, as well as an equivalent example for the second ($r=2$) hierarchy state at $\nu=2/5$ filling, is shown in Fig.~\ref{fig:example_geom}. 

\begin{table}
	\centering
	$\nu=1/3$ state
	\vspace{0.5em}
	\begin{ruledtabular}
		\begin{tabular}{c c c c c c c c c c}
			$p$ & $l_x$ & $l_y$ & $L_x$ & $L_y$ & $N$ & $N_u$ & $N_s$ & $N_o$ & $\left.\Delta/W\right|_{\kappa=0.3}$ \\
			\hline
			1 & 3 & 1 & 1 & 6 & 2 & 18 & 36 & 72 & \multirow{2}{*}{3.64} \\
			1 & 3 & 1 & 1 & 9 & 3 & 27 & 54 & 108 \\
			1 & 4 & 1 & 1 & 6 & 2 & 24 & 48 & 96 & \multirow{2}{*}{2.10} \\
			1 & 4 & 1 & 1 & 9 & 3 & 36 & 72 & 144 \\
			1 & 5 & 1 & 1 & 6 & 2 & 30 & 60 & 120 & \multirow{2}{*}{77.6} \\
			1 & 5 & 1 & 1 & 9 & 3 & 45 & 90 & 180 \\
		\end{tabular}
	\end{ruledtabular}
	\centering
	\vspace{1em}
	$\nu=2/5$ state
	\vspace{0.5em}
	\begin{ruledtabular}
		\begin{tabular}{c c c c c c c c c c}
			$p$ & $l_x$ & $l_y$ & $L_x$ & $L_y$ & $N$ & $N_u$ & $N_s$ & $N_o$ & $\left.\Delta/W\right|_{\kappa=0.3}$ \\
			\hline
			1 & 3 & 1 & 1 & 5 & 2 & 15 & 30 & 60 & \multirow{2}{*}{3.64} \\
			1 & 3 & 1 & 1 & 10 & 4 & 30 & 60 & 120 \\
			1 & 4 & 1 & 1 & 5 & 2 & 20 & 40 & 80 & \multirow{2}{*}{2.10} \\
			1 & 4 & 1 & 1 & 10 & 4 & 40 & 80 & 160 \\
			1 & 5 & 1 & 1 & 5 & 2 & 25 & 50 & 100 & \multirow{2}{*}{77.6} \\
			1 & 5 & 1 & 1 & 10 & 4 & 50 & 100 & 200 \\
		\end{tabular}
	\end{ruledtabular}
	\caption{\label{tab:lat_geom}A selection of lattice geometries to realize the $\nu=1/3$ and $2/5$ FQH states in the moir{\'e} Hamiltonian. Since we are working in Landau gauge in the $x$ direction, $l_y=1$ is fixed. The number of lattice unit cells is given as $N_u=L_x l_x \times L_y l_y$, with the number of lattice sites $N_s=2N_u$ and the number of orbital sites $N_o=2N_s$. The desired filling factor fixes the number of particles such that $\nu=n/n_\phi=(N/N_u)/(p/q)$, where $q=l_x$ in our choice of gauge. $\Delta/W$ denotes the gap-to-width ratio for the lowest-energy band pairs for Eq.~(\ref{eq:single_ham}) with $\kappa=0.3$.}
\end{table}  

This simple case study highlights the typical considerations for selecting a lattice configuration at a given filling factor. Since the computations scale exponentially in the cylinder circumference, we are additionally restricted in our choice of $L_y$. Table~\ref{tab:lat_geom} shows a selection of the simplest numerically accessible lattice geometries for realizing the fermionic Laughlin and $r=2$ hierarchy states. The table lists the prototypical geometries with $p=1$, known to be robust in the Hofstadter model and motivated by the flux densities for topological flat bands in Sec.~\ref{subsec:top_flat_bands}. We further tabulate the gap-to-width ratios for the lowest-energy bands in Eq.~(\ref{eq:single_ham}) and note that, although increasing $q$ decreases the band widths, it does not necessarily increase the flatness ratio.

\subsection{Integer and fractional quantum Hall states}
\label{subsec:FCIs}

In order to produce an integer quantum Hall (IQH) state, we need to fill the lowest band (or any integer multiple of bands) with non-interacting particles. For particles filling the lowest $m$ bands in the spectrum, this yields a Hall conductivity of $\sigma_\text{H}=e^2/h \sum_i C_i$, where $C_i$ is the Chern number of band $i$. In order to produce a FQH state, however, we need to fractionally fill isolated flat bands with interacting particles. For particles filling such bands with appropriate filling fraction $\nu$, the Hall conductivity is $\sigma_\text{H}=(e^2/h)C\nu$~\cite{Grushin15, Schoonder19}. For the demonstration of FQH states in this section, we choose the two-particle $\nu=1/3$ and $2/5$ states both at flux density $n_\phi=1/3$, as detailed in Table~\ref{tab:lat_geom}. For the demonstration of a IQH state, we take the corresponding non-interacting system at unit filling. Note that in these simulations we fill the bands for the $\nu=1/3$ and $2/5$ states corresponding to the honeycomb Hofstadter model, and then tune $\kappa\in[0, 1]$. The characterization of topological order in this section is based on the works of Grushin~\textit{et~al.}~\cite{Grushin15} and Cincio and Vidal~\cite{Cincio13}. 

\begin{figure}
	\centering{$\nu=1/3$ state}\\
	\includegraphics[width=\linewidth]{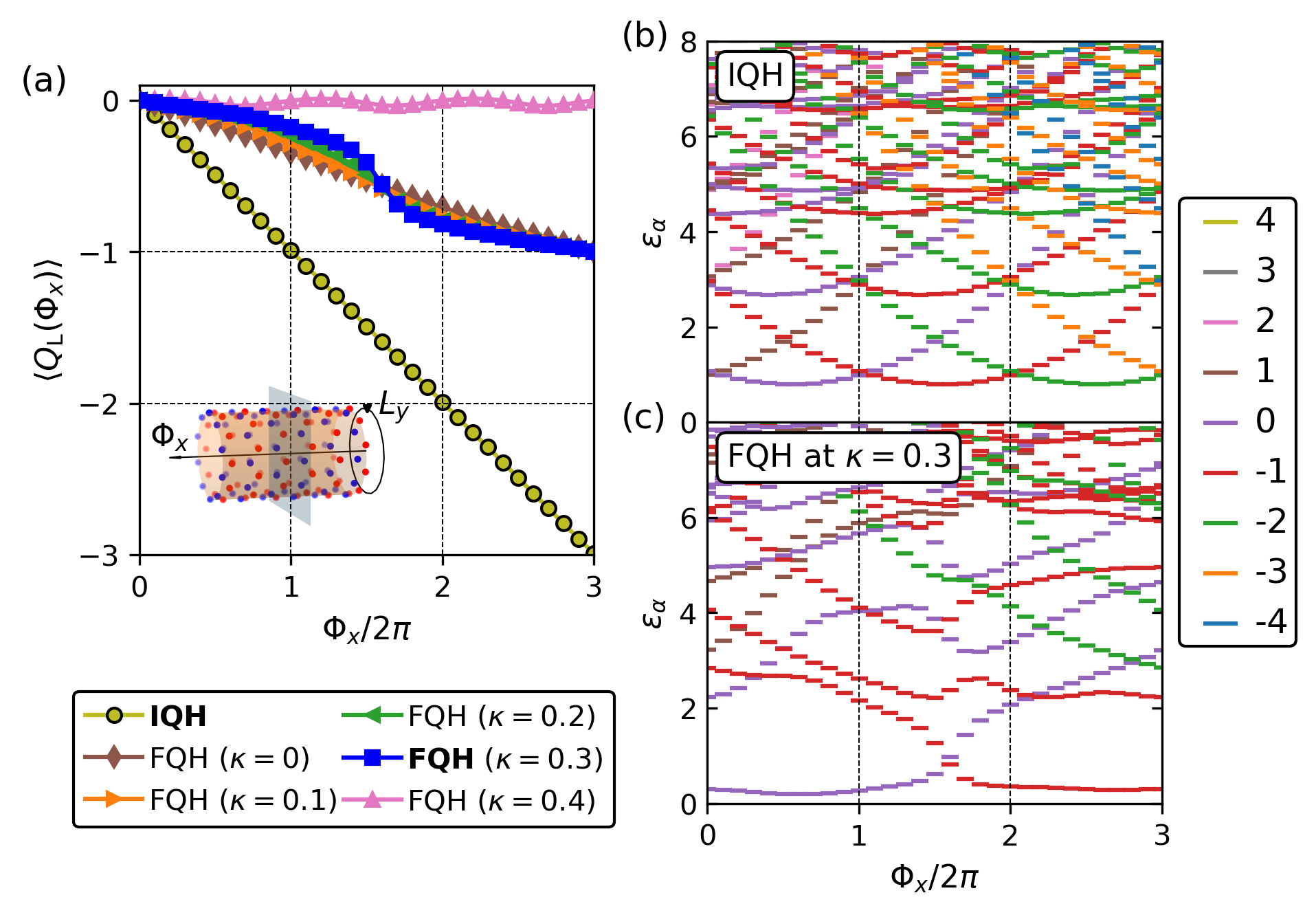}
	\centering{$\nu=2/5$ state}\\
	\includegraphics[width=\linewidth]{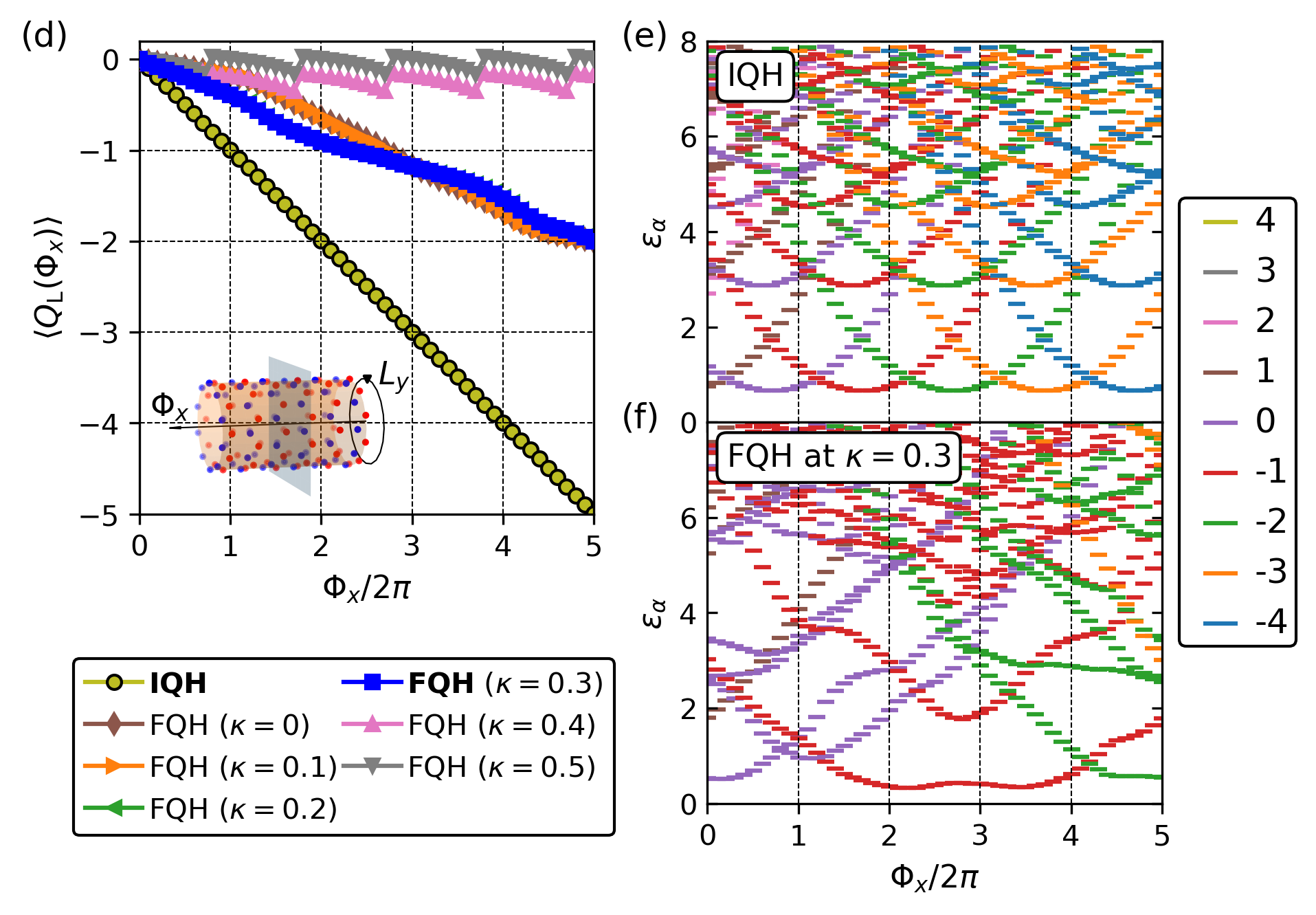}
	\caption{\label{fig:phi_flow} IQH and FQH flux insertion examples from many-body calculations of Eq.~(\ref{eq:single_ham}) at $n_\phi=1/3$ with [(a)--(c)] $\nu=1/3$ and [(d)--(f)] $\nu=2/5$, using iDMRG on a thin cylinder. For the IQH examples, the simulation parameters are $\nu=1$ and $U=V=0$, whereas for the FQH examples we use $U=10V=100 t_1$ and $0\leq\kappa\leq0.5$. The simulations were performed with a circumference of $L_y=6$ for $\nu=1/3$ and $L_y=5$ for $\nu=2/5$, at a bond dimension of $\chi=150$. [(a),(d)]~Charge pumping as we thread a flux $\Phi_x$ through the center of the cylinder (sketch inset). The charge pumping for the FQH state is plotted for a range of $\kappa$, as indicated in the legend. [(b),(c),(e),(f)]~Entanglement spectrum flow for the $(k_x, k_y)=(0,0)$ momentum sector, for the states marked in bold in [(a),(d)], where the energy levels are labeled by the $U(1)$ charge sector. Note that the full entanglement spectra are not shown, in order to emphasize the flow of the low-lying states.}
\end{figure}

In Fig.~\ref{fig:phi_flow} we show the behavior of the IQH and FQH states under flux insertion~\cite{Gong14, Zhu16_1, Zhu16_2}. Figure~\ref{fig:phi_flow}(a) shows the expected charge on the left half of the cylinder, $\braket{Q_\text{L}}$, as a function of the flux threaded through the cylinder, $\Phi_x$. A sketch of the cylinder geometry is shown in the insert. As can be seen for the IQH case, after one flux quantum is threaded through the cylinder, exactly one unit of charge is pumped across the cut~\cite{Laughlin83}. This result verifies the complete filling of a $C=1$ band. For the FQH state, we notice that now one unit of charge is pumped across the cut after an insertion of three flux quanta. This corresponds to a Hall conductivity of $\sigma_\text{H}=e^2/3h$, which is the $\nu=1/3$ filling of a $C=1$ band. Additionally, we plot the charge pumping result for a range of $\kappa$, as indicated in the legend. These data show that, at $\kappa=0$, we obtain a linear charge pumping relation, whereas when the tuning parameter is increased the curve is smoothly distorted to a discontinuity. Eventually, for $\kappa>0.3$ the charge pumping goes to zero -- signaling a transition to an insulating phase.

In Fig.~\ref{fig:phi_flow}(b,c) we show the corresponding entanglement spectral flow~\cite{Zhu15_1, Zhu15_2}. Since the Schmidt values are positive and monotonic, the entanglement energies are typically defined $\epsilon_\alpha \equiv -\log \Lambda_\alpha^2$~\cite{Li08}. Hence, the larger the bond dimension, the more entanglement energies in the spectrum. Figure~\ref{fig:phi_flow}(b) shows the entanglement spectrum as a function of flux threaded through the cylinder in the IQH case. In the plot, the energy levels are labeled by the corresponding charge quantum numbers, arising from the $U(1)$ charge conservation symmetry of the Hamiltonian. Here, it can be seen that the entanglement spectrum is symmetric under multiples of one unit of flux threaded through the cylinder, up to charge sector labeling~\cite{Alexandra11}. The lowest-energy state at $\Phi_x=0$, for example, starts in charge sector 1 and then shifts by one after each multiple of flux insertion corresponding to $\sigma_\mathrm{H}\propto C\nu=1$. For the FQH case, we see analogous behavior after multiples of three flux quanta, again in agreement with the charge pumping result. Specifically, in Fig.~\ref{fig:phi_flow}(c) we notice that the sharp flow of states into one another at $\Phi_x\sim3\pi$ is analogous to the behavior of the charge pumping curves in Fig.~\ref{fig:phi_flow}(a) as we approach the transition.

Furthermore, in Fig.~\ref{fig:phi_flow}(d,e,f) we show the flux insertion behavior for the $\nu=2/5$ state. In Fig.~\ref{fig:phi_flow}(d) we show analogous results for the charge pumping. For the FQH state, we see that now two charges are pumped across the cut after the insertion of five flux quanta, in agreement with the expected Hall conductivity $\sigma_\mathrm{H}\propto C\nu$ for a $C=1$ band. Plotting this charge pumping as a function of $\kappa$, we now see that the charge pumping breaks abruptly, not showing a smooth distortion as in Fig.~\ref{fig:phi_flow}(a). Furthermore, at values of $\kappa>0.3$ the charge pumping does not immediately go to zero but rather gradually tends to zero in the interval $0.3<\kappa<0.5$ indicating a transient `metallic' phase prior to the insulating phase. We note that since this region is diminished as the bond dimension is increased this may be an artifact of the numerics. The spectral flow concurs with the charge pumping, as for the Laughlin state. For the FQH state, the entanglement spectrum is now symmetric every five flux quanta with the charge sector labeling shifting by two each time. 

\begin{figure}
	\includegraphics[width=\linewidth]{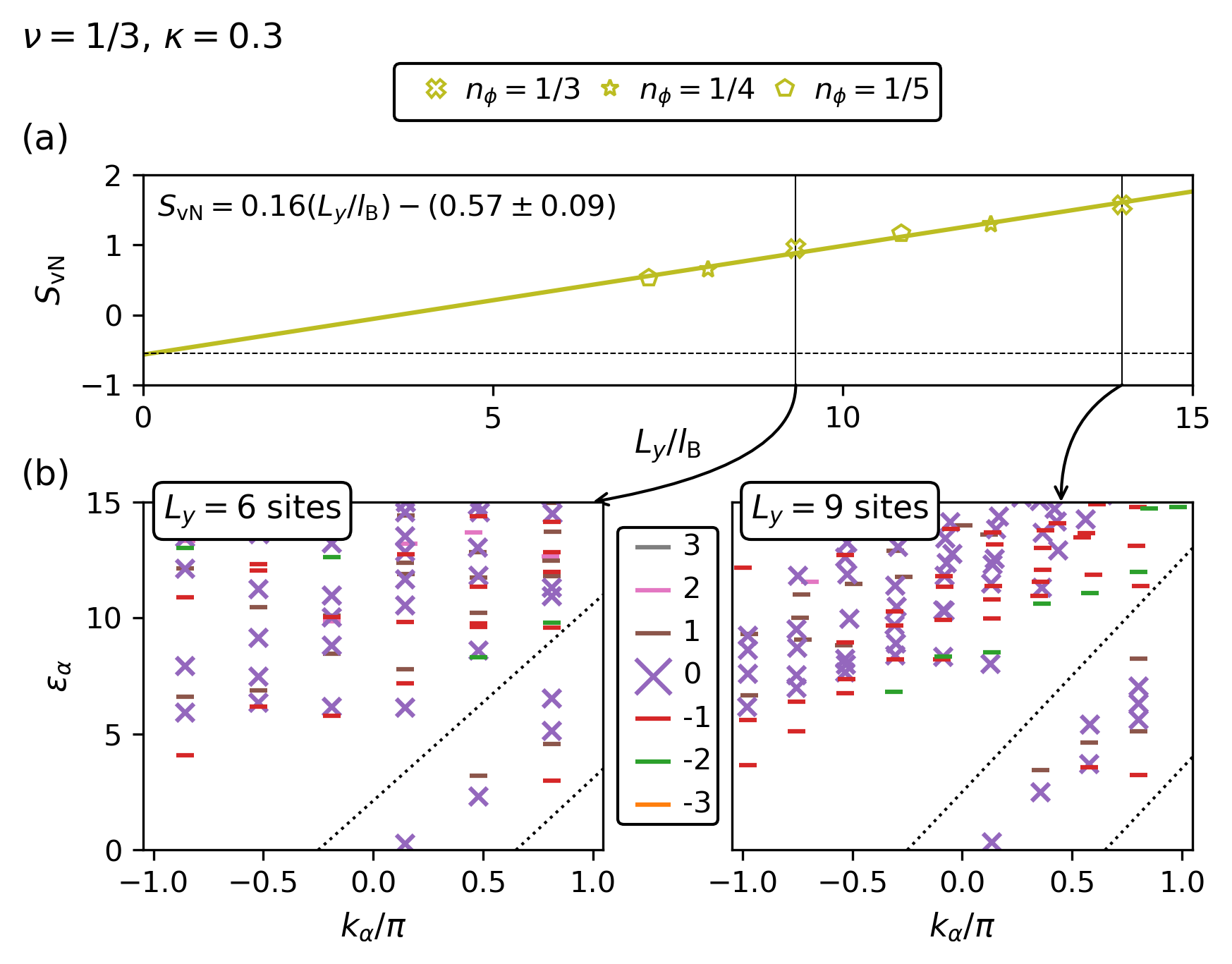}\\
	\caption{\label{fig:Ly_flow} FQH entropy scaling example for the $\nu=1/3$ state defined in Fig.~\ref{fig:phi_flow}. (a)~Scaling of the von Neumann entanglement entropy from bisecting the cylinder, $S_\mathrm{vN}$, with cylinder circumference in units of the magnetic length $L_y/l_\text{B}$. The entropy values are obtained by extrapolating data up to bond dimension $\chi=300$, and the $\gamma=0.549$ asymptote for the $\nu=1/3$ Laughlin state is marked with a dashed line. (b)~Momentum-resolved entanglement spectra for the FQH state with $n_\phi=1/3$ at $L_y=6$ and $9$ sites, up to a rotation in momentum space. The state for $L_y=9$ is calculated at a higher bond dimension of $\chi=500$. The energy levels are labeled by the $U(1)$ charge sector, and the dashed lines guide the edge states. Note that the full entanglement spectra are not shown, in order to highlight the edge state counting.}
\end{figure}

We further support our claim of FQH states through the use of entanglement scaling~\cite{Grushin15, Schoonder19}. In Fig.~\ref{fig:Ly_flow}(a) we show the scaling of the von Neumann entropy, $S_\mathrm{vN}$, as a function of cylinder circumference. Here we follow the approach of Schoonderwoerd~\textit{et~al.}~\cite{Schoonder19} and plot the cylinder circumference $L_y$ in units of magnetic length $l_\text{B}=\sqrt{\hbar/eB}\sim n_\phi^{-1/2}$ as an inexpensive way to add points for different flux densities. It is well known from the `area law' of entanglement that the entanglement entropy scales as $S=c L_y - \gamma_i$, where $c$ is a non-universal constant dependent on microscopic system parameters, and $\gamma_i$ is the topological contribution dependent on the quantum dimensions of potential quasi-particles supported by the system~\cite{Kitaev06,Levin06}. In particular, for the $\nu=1/3$ Laughlin state, the topological entanglement entropy is known to be $\gamma=0.549$, which is in agreement with the computed value $\gamma=0.57\pm0.09$. The data points in the Fig.~\ref{fig:Ly_flow}(a) were obtained by extrapolating the entropy in the $\chi\to\infty$ limit based on data up to $\chi=300$, as discussed in Appendix~\ref{sec:iDMRG}. We note that, in contrast to the charge pumping in Fig.~\ref{fig:phi_flow}, a large bond dimension is required for the entropy scaling to obtain accurate results. We do not show the scaling plot for the $\nu=2/5$ state as it did not converge adequately for the bond dimensions considered. However, preliminary results point to a topological entanglement in agreement with Abelian conformal field theory (CFT) predication of $\gamma=0.80$~\cite{Estienne15}. We intend to investigate this further in a follow-up work.

In Fig.~\ref{fig:Ly_flow}(b,c) we present the (azimuthal) momentum-resolved entanglement spectra for the two system circumferences at $n_\phi=1/3$. In both cases, we observe $L_y$ momentum sectors with the CFT edge state counting $\{1, 1, 2, 3, \dots\}$ for each charge sector, consistent with the $\nu=1/3$ Laughlin state~\cite{Li08, Liu13}. Note that as the cylinder circumference is increased the gaps in the spectrum become more pronounced and the edge counting can be verified to higher order, since finite-size effects are suppressed.

\begin{figure}
	\includegraphics[width=\linewidth]{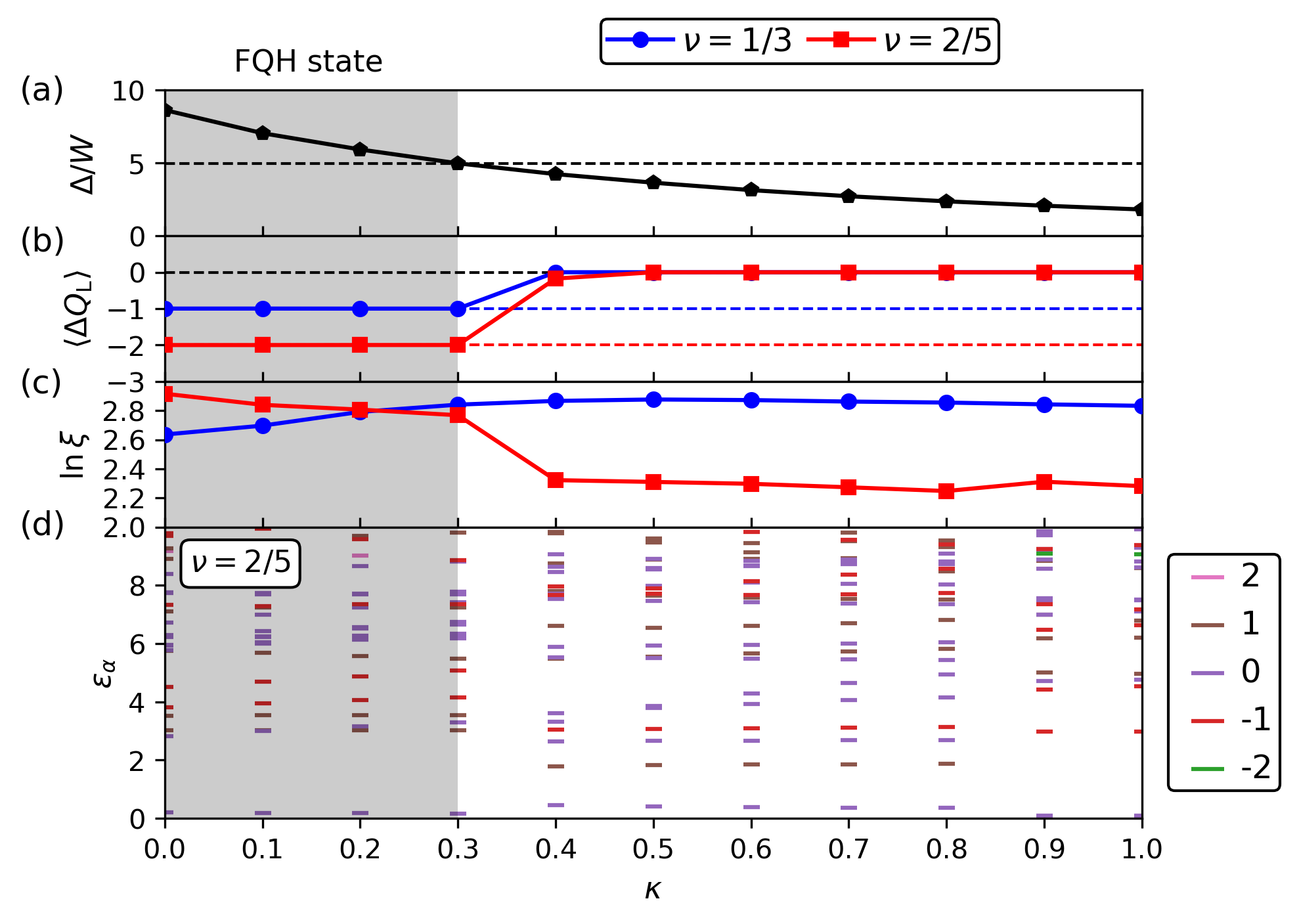}\\
	\caption{\label{fig:kappa_flow} FQH to topologically trivial phase transition with respect to the dimensionless tuning parameter, $\kappa$. Parameters are defined in Fig.~\ref{fig:phi_flow} with a bond dimension of $\chi=200$. (a) Gap-to-width ratio, $\Delta/W$, of the lowest bands. (b) Average charge on the left side of the cylinder, $\braket{Q_\mathrm{L}}$, after a flux insertion of $\Delta\Phi_x = 6\pi$ for the $\nu=1/3$ state and $\Delta\Phi_x = 10\pi$ for the $\nu=2/5$ state. (c) Correlation length, $\xi$. (d) Entanglement spectra for the $(k_x,k_y)=(0,0)$ momentum sector, plotted as a function of $\kappa$ for the $n_\phi=1/3$ system at $\nu=2/5$ filling with circumference $L_y=5$. The value of $\kappa=0.3$, discussed throughout this paper, is marked by the boundary of the phase transition. As before, the energy levels are labeled by their $U(1)$ charge sector.}
\end{figure}

Finally, in Fig.~\ref{fig:kappa_flow} we present results demonstrating the phase transition from FQH to topologically trivial phases with tuning parameter $\kappa$. In Fig.~\ref{fig:kappa_flow}(a), we show the reduction in gap-to-width ratio as $\kappa$ is increased. This trend indicates a weakening of the gap condition discussed in Sec.~\ref{subsec:many_ham}. In Fig.~\ref{fig:kappa_flow}(b), we show the average charge pumped across the system after three (five) flux insertions for the $\nu=1/3$ ($\nu=2/5$) state, corresponding to the FQH results in Fig.~\ref{fig:phi_flow}(a,d). We can see that the FQH phase survives as high as $\kappa=0.3$, which motivates the value chosen throughout this paper. This choice also maintains the relative orders of magnitude of hopping parameters $t_1\sim 10^2 |t_2| \sim 10 t_2'$ from the literature~\cite{Yuan18,ErratumYuan18,Koshino18}. We further observe that the Laughlin state has a direct transition to an insulating phase, whereas the charge pumped for the $\nu=2/5$ state does not go directly to zero, indicating a transient metallic phase at $\kappa=0.4$, possibly as a result of using an insufficient bond dimension. In Fig.~\ref{fig:kappa_flow}(c), we see the nature of the two transitions reflected in the correlation length plot. For the $1/3$ state, we observe a gradual saturation, whereas for the $2/5$ state we see a discontinuous jump. This reflects the transitions observed from the flux insertion analysis. Finally, in Fig.~\ref{fig:kappa_flow}(d) we highlight this discontinuous transition for the $2/5$ state by examining the low-lying states of the entanglement spectrum. From this analysis we conclude that the orbital-polarized FQH states observed at $\nu=1/3$ and $2/5$ are suppressed as the orbital/band mixing is increased and we demonstrate that these states can persist remarkably up to the same order of magnitude as typical hopping parameters for the moir{\'e} Hamiltonian.

\section{Discussion and Conclusions}
\label{sec:conclusion}

In this paper, we have presented evidence for $\nu=1/3$ and $2/5$ FQH states for the many-body Hamiltonian in Eq.~(\ref{eq:many_ham}) at $\kappa=0.3$. Subsequently, we revealed that $\kappa=0.3$ is in fact the largest tuning value for the hopping amplitudes before the state undergoes a transition, for these lattice configurations. The findings can be summarized into three main results. First, we found evidence of FQH states for this moir{\'e} superlattice Hamiltonian. Second, we tuned from the dominant term to the full Hamiltonian to show that these states persist up to $\approx30\%$ of the typical hopping parameters, after which they undergo a transition into an insulating phase. Third, we demonstrate that the $\nu=1/3$ and $2/5$ states for these configurations are a feature of the underlying honeycomb Hofstadter model and are consequently suppressed as we tune to the full model.

From the findings presented in Sec.~\ref{sec:results}, the first two of these conclusions are clear, but the third requires some extra discussion. When we consider our Hamiltonian in Eq.~(\ref{eq:many_ham}) at $\kappa=0$, we essentially have two identical non-interacting copies of the honeycomb Hofstadter model, one corresponding to $p_x$ orbitals and the other to $p_y$. The lowest band for the single-orbital honeycomb Hofstadter model has Chern number $C=1$. As we increase the tuning parameter $\kappa$, the orbital mixing means that these two lowest $C=1$ bands hybridize to a $C=2$ band. Therefore, when we fractionally fill a $C=1$ band and stabilize a FQH state for the honeycomb Hofstadter model and gradually tune to the full effective model, these FQH states are destabilized. To analyze this in more detail, we examine the two states in turn. First for the $\nu=1/3$ state at $n_\phi=1/3$, we need to fill one of the 12 bands up to $1/3$. As mentioned before, for fermions filling a $C=1$ band this is a valid FQH state in accordance with the hierarchy in Eq~(\ref{eq:Moller}). However, once the lowest two degenerate bands have hybridized, this would correspond to filling a $C=2$ band up to $1/6$. This is not a valid FQH state and so as $\kappa$ is increased this state is destabilized. Second for the $\nu=2/5$ state at $n_\phi=1/3$, we need to fill one of the 12 bands up to $2/5$. As before, this yields a valid state in accordance with hierarchy. However, now when the lowest orbital-degenerate band pair hybridizes, this would correspond to filling a $C=2$ band up to $1/5$. Even though this is still a valid FQH state, it is not stabilized by the system parameters. The conclusion that these states are orbital polarized is confirmed by the significantly faster convergence of the iDMRG algorithm with an orbital-polarized initial product state. Furthermore, this distinction between the breakdown for the $\nu=1/3$ and $2/5$ states may account for the different nature of the transitions observed in Figs.~\ref{fig:phi_flow}(a,d) and~\ref{fig:kappa_flow}(c).        

Throughout this paper, we have focused on the extent to which the orbital-polarized $\nu=1/3$ and $2/5$ states survive in the moir{\'e} superlattice Hamiltonian. At this stage, it is important to note that we do not exclude the possibility of realizing these states at $\kappa>0.3$ in general, since there are many ways of realizing the states for different flux densities. For example, we have observed that the band structure for Eq.~(\ref{eq:single_ham}) at $n_\phi=10/11$ yields particularly flat bands at low energies, but is not tractable for our many-body calculations at the moment. For the lattice geometries in Sec.~\ref{subsec:lat_geom}, we instead study configurations with $p=1$ due to computational expense, and because these configurations are known to be particularly stable in the Hofstadter model. With improvements to the algorithm~\cite{Motruk16, Kantian19, Milstead19}, it would also be interesting to investigate how long the states can survive at other flux densities, as well as the presence of potential competing states at larger MPS unit-cell dimensions.

The research questions in this paper can also be extended in other directions, such as examining the persistence of underlying fractional states for lattice generalizations of FQH states (fractional Chern insulators~\cite{Regnault11,Bergholtz13}), further non-primary composite fermion states (with $|r|>2$)~\cite{Laeuchli13} or non-Abelian states~\cite{Liu13}, as well as the stability of such states in the continuum limit~\cite{Andrews18}. In the longer term, one could interface \textit{ab-initio} calculations (using appropriate Wannierizations) to obtain effective Hamiltonians customized to realistic twisted bilayer or multi-layer structures, or examine the bulk topological invariants for systems with specific crystallographic symmetries~\cite{Chen12}. In all cases, we emphasize the persistence of fractional states from underlying Hamiltonians. We expect that it will be possible to exploit this knowledge in the future and comment on the fractional states present in complicated moir{\'e} Hamiltonians in certain regimes simply by looking at their dominant terms.  

\begin{acknowledgments}
This work is in memory of Alexey Soluyanov, who passed away during the preparation of this paper. As a mentor and as a friend, his guidance and support continue to live on through everyone who knew him. 

B.~A. thanks Titus Neupert and Mark Fischer for critical reading of the paper, and also acknowledges useful discussions with Johannes Hauschild, Leon Schoonderwoerd, Arkadiy Davydov, Gunnar M{\"o}ller, Adolfo Grushin, Johannes Motruk, Tobias Wolf, and Ashvin Vishwanath. Especially, B.~A. thanks Johannes Hauschild for help and advice in using the \textsc{TeNPy} library~\cite{tenpy} and with the error analysis in Appendix~\ref{sec:iDMRG}. This project was funded by the Swiss National Science Foundation under Grant No. PP00P2\_176877.
\end{acknowledgments}

\onecolumngrid
\appendix

\section{Details of the Peierls substitution}
\label{sec:peierls}

As mentioned in Sec.~\ref{subsec:single_ham}, we work in the Landau gauge in the $x$ direction such that our perpendicular magnetic field is described by the vector potential $\mathbf{A}=Bx\hat{\mathbf{e}}_y$. The Peierls phases acquired by hopping from site $i\equiv(X_i,Y_i)$ to site $j\equiv (X_j,Y_j)$ are given as $\theta_{ij}=(2\pi/\phi_0)\int_i^j \mathbf{A}\cdot \mathrm{d}\mathbf{l}$, where $\phi_0$ is the flux quantum and $\mathrm{d}\mathbf{l}=(\mathrm{d}x,\mathrm{d}y)$ is an infinitesimal line element. By making an appropriate parametrization, for example
\begin{equation}
\begin{cases}
x = X_i + (X_j - X_i)\tau \\
y = Y_i + (Y_j - Y_i)\tau,
\end{cases}
\end{equation}
where $\tau\in[0,1)$, the Peierls phases can be written as
\begin{equation}
\theta_{ij}=\left(\frac{2\pi B}{\phi_0}\right)(Y_j-Y_i) \left(X_i + \frac{X_j - X_i}{2} \right).
\end{equation}
From this, it can be seen that the Peierls phases depend on \emph{absolute} $x$ coordinates but only \emph{relative} $y$ coordinates.

The Peierls phases acquired for nearest-neighbor hoppings on a honeycomb lattice are already well known~\cite{Rammal85}. In this case, using the definitions of Fig.~\ref{fig:lattice_bz}(a), we may write the Peierls phases for a (charged) particle at coordinates $(n,m)$ as
\begin{equation}
\label{eq:phases1}
\theta_{ij}^{(m,n)}=\begin{cases}
- 2\pi n_\phi m/3 & j=(m,n- 2) \\
+ \pi n_\phi (m+1/2)/3 & j=(m+1, n+ 1) \\
+ \pi n_\phi (m-1/2)/3 & j=(m-1, n+ 1)
\end{cases},
\end{equation}
where $n_\phi\equiv 2\sqrt{3}Bb^2/\phi_0$ is the magnetic flux per unit cell (flux density). This allows us to write a tight-binding equation for a particle at $(m,n)$, as well as for its three nearest neighbors. Using these four simultaneous equations, we can solve for the A sublattice. Finally, invoking the plane-wave ansatz in the $y$ direction, due to our choice of gauge, leaves us with
\begin{equation}
\label{eq:hex1}
(E^2-3)\psi_{m} = C' \psi_{m-2} + B'^*_{m-1} \psi_{m-1} + B'_m \psi_{m+1} + C'^* \psi_{m+2},
\end{equation}
where
\begin{align*}
B'_m &= 2 e^{\mathrm{i}\pi n_\phi/3} \cos (\pi n_\phi (m+1/2) + 3 k_y c), \\
C' &= e^{\mathrm{i}\pi n_\phi/3}.
\end{align*}

The Peierls phases for second-nearest-neighbor hopping on a triangular lattice have also been previously investigated~\cite{Oh00}. This analysis only requires slight modification for fifth-nearest-neighbor hopping on the honeycomb lattice, to account for the presence of A and B sublattices. To this end, we follow the same procedure as above. Using the definitions of Fig.~\ref{fig:lattice_bz}(a), we may write the Peierls phases for a (charged) particle at coordinates $(m,n)$ as
\begin{equation}
\label{eq:hex5_peierls}
\theta_{ij}^{(m,n)}=\begin{cases}
+ 2\pi n_\phi m & j=(m, n+ 6) \\
- \pi n_\phi(m + 3/2) & j=(m+3,n- 3) \\
- \pi n_\phi(m-3/2) & j=(m-3, n- 3) \\
\hline
- 2\pi n_\phi m & j=(m, n- 6) \\
+ \pi n_\phi(m + 3/2) & j=(m+3,n+ 3) \\
+ \pi n_\phi(m-3/2) & j=(m-3, n+ 3)
\end{cases},
\end{equation}
where the horizontal line separates the contributions from the A and B triangular sublattices. We then write a tight-binding equation for the particle at $(m,n)$ as well as for its fifth-nearest neighbors. We solve these seven simultaneous equations for the A sublattice, and invoke the plane-wave ansatz in the $y$ direction to yield:
\begin{equation}
\label{eq:hex5}
(E^2-3)\psi_m = G'_{m-4}\psi_{m-6} + D'_{m-2}\psi_{m-3} + A'_m\psi_{m} + D'_{m+1}\psi_{m+3} + G'_{m+2}\psi_{m+6},
\end{equation}
where
\begin{align*}
A'_m &= 2\cos(4\pi n_\phi m + 12 k_y c) + 2\cos(2\pi n_\phi(m-3/2)+6 k_y c) + 2\cos(2\pi n_\phi(m+3/2)+6k_y c)+3, \\
D'_m &= 2\cos(\pi n_\phi(m-5/2)+3k_y c) + 2\cos(\pi n_\phi(m+7/2)+3k_y c) \\
&\hphantom{=} + 2\cos(3\pi n_\phi(m-1/2)+9k_y c) +2\cos(3\pi n_\phi(m+3/2)+9k_y c), \\
G'_m &= 2 \cos(2\pi n_\phi(m+1)+6k_y c) + 2\cos(3\pi n_\phi).
\end{align*}

At this stage, it is possible to combine Eqs.~(\ref{eq:hex1}) and~(\ref{eq:hex5}) in an appropriate superposition to account for the $t_1$ and $t_2$ terms in Eq.~(\ref{eq:single_ham}). As a last step, we analyze the effect of the orbital mixing ($t_2'$) term in the Hamiltonian. Written out explicitly, this contribution is given as
\begin{equation}
H_{t_2'} = t_2' \left[ \sum_{\braket{ij}_5} c_{x,i}^\dagger c_{y, j} - (x \leftrightarrow y) + \text{H.c.} \right].
\end{equation}
Here the honeycomb lattice not only has A and B sites, but additionally each site has $p_x$ and $p_y$ orbitals. In this tight-binding approximation, we model the orbitals to lie on the same site. Hence, the Peierls phases are the same as in Eq.~(\ref{eq:hex5_peierls}) but now the $p_x\to p_y$ hoppings come with a minus sign. Following the standard procedure, we find that the bands for both orbitals are equivalent, with eigenenergies given by
\begin{equation}
(E^2-3)\psi_m = G''_{m-4}\psi_{m-6} + D''_{m-2}\psi_{m-3} + A''_m\psi_{m} + D''_{m+1}\psi_{m+3} + G''_{m+2}\psi_{m+6},
\end{equation}
where
\begin{align*}
A''_m &= -2\cos(4\pi n_\phi m + 12 k_y c) - 2\cos(2\pi n_\phi(m-3/2)+6 k_y c) - 2\cos(2\pi n_\phi(m+3/2)+6k_y c)-9, \\
D''_m &= -2\cos(\pi n_\phi(m-5/2)+3k_y c) - 2\cos(\pi n_\phi(m+7/2)+3k_y c) \\
&\hphantom{=}- 2\cos(3\pi n_\phi(m-1/2)+9k_y c) -2\cos(3\pi n_\phi(m+3/2)+9k_y c), \\
G''_m &= -2 \cos(2\pi n_\phi(m+1)+6k_y c) - 2\cos(3\pi n_\phi).
\end{align*}

Writing the flux per unit cell as a rational fraction $n_\phi=p/q$, we find that the Bloch condition in the $x$ direction is $\psi_{m+M}=e^{\mathrm{i} k_x M b}\psi_m$ with $M=q$ $(2q)$ for even (odd) $p$, and $1 \leq m \leq M$~\cite{Rammal85}. Hence, the complete $M\times M$ single-particle Hamiltonian matrix corresponding to Eq.~(\ref{eq:single_ham}) may be written as
\begin{equation}
\label{eq:matrix_twist}
\mathbf{H} =\begin{pmatrix}
A_1 & B_1 & C^* & D_2 & 0 & 0 & G_3 & \dots & 0 & \tilde{G}_{M-3} & 0 & 0 & \tilde{D}_{M-1} & \tilde{C} & \tilde{B}^*_M \\
B^*_1 & A_2 & B_2 & C^* & D_3 & 0 & 0 & \dots & 0 & 0 & \tilde{G}_{M-2} & 0 & 0 & \tilde{D}_{M} & \tilde{C} \\
C & B^*_2 & A_3 & B_3 & C^* & D_4 & 0 & \dots & 0 & 0 & 0 & \tilde{G}_{M-1} & 0 & 0 & \tilde{D}_1 \\
D_2 & C & B^*_3 & A_4 & B_4 & C^* & D_5 & \dots & 0 & 0 & 0 & 0 & \tilde{G}_M & 0 & 0 \\
0 & D_3 & C & B^*_4 & A_5 & B_5 & C^* & \dots & 0 & 0 & 0 & 0 & 0 & \tilde{G}_1 & 0 \\
0 & 0 & D_4 & C & B^*_5 & A_6 & B_6 & \dots & 0 & 0 & 0 & 0 & 0 & 0 & \tilde{G}_2 \\
G_3 & 0 & 0 & D_5 & C & B^*_6 & A_7 & \dots & 0 & 0 & 0 & 0 & 0 & 0 & 0 \\
\vdots & \vdots & \vdots & \vdots & \vdots & \vdots & \vdots & \ddots & \vdots & \vdots & \vdots & \vdots & \vdots & \vdots & \vdots \\
0 & 0 & 0 & 0 & 0 & 0 & 0 & \dots & A_{M-6} & B_{M-6} & C^* & D_{M-6} & 0 & 0 & G_{M-6} \\
\bar{G}_{M-3} & 0 & 0 & 0 & 0 & 0 & 0 & \dots & B^*_{M-6} & A_{M-5} & B_{M-5} & C^* & D_{M-5} & 0 & 0 \\
0 & \bar{G}_{M-2} & 0 & 0 & 0 & 0 & 0 & \dots & C & B^*_{M-5} & A_{M-4} & B_{M-4} & C^* & D_{M-4} & 0 \\
0 & 0 & \bar{G}_{M-1} & 0 & 0 & 0 & 0 & \dots & D_{M-6} & C & B^*_{M-4} & A_{M-3} & B_{M-3} & C^* & D_{M-3} \\
\bar{D}_{M-1} & 0 & 0 & \bar{G}_{M} & 0 & 0 & 0 & \dots & 0 & D_{M-5} & C & B^*_{M-3} & A_{M-2} & B_{M-2} & C^* \\
\bar{C}^* & \bar{D}_{M} & 0 & 0 & \bar{G}_1 & 0 & 0 & \dots & 0 & 0 & D_{M-4} & C & B^*_{M-2} & A_{M-1} & B_{M-1}\\
\bar{B}_{M} & \bar{C}^* & \bar{D}_1 & 0 & 0 & \bar{G}_2 & 0 & \dots & G_{M-6} & 0 & 0 & D_{M-3} & C & B^*_{M-1} & A_{M}
\end{pmatrix}, 
\end{equation}
where $A_m=t_2 A'_m + t_2' A''_m$, $B_m=t_1 B'_m$, $C=t_1 C'$, $D_m = t_2 D'_m + t_2' D''_m$, $G_m=t_2 G'_m + t_2' G''_m$, and $\kappa\equiv 1$. We additionally define the shorthand $\bar{A}\equiv A e^{\mathrm{i}\delta}$, $\tilde{A}\equiv A e^{-\mathrm{i}\delta}$, and $\delta\equiv k_x M a /2$. Note that there are $2M$ eigenenergies given by $E_i=\pm\sqrt{\lambda_i+3}:i\in[0,M)$, due to the presence of the A and B sublattices. Sweeping over flux per unit cell (at $\mathbf{k}=\mathbf{0}$), yields the single-particle energy spectra in Fig.~\ref{fig:butterfly}.

\section{Details of the iDMRG simulation}
\label{sec:iDMRG}

Due to the $p_x$ and $p_y$ orbitals and the fifth-nearest-neighbor hopping terms in Eq.~\ref{eq:many_ham}, the iDMRG simulations in this paper are demanding. To quantify this, we examine the (upper estimate of) algorithm scaling for the runtime of a single bond update $\sim O(\chi^3 D d^3 + \chi^2 D^2 d^2)$ and for the memory usage $\sim O(\chi^2 d N + 2 \chi^2 D N)$, where $\chi$ is the MPS bond dimension, $D$ is the maximum MPO bond dimension, $d$ is the single-site Hilbert-space dimension, and $N$ is the total number of sites (including orbital sites) in the MPS unit cell. In this paper, the MPS unit cell has dimensions $(L_x,L_y)$, where $L_x$ is in units of magnetic unit-cell width. For the single-orbital fermionic honeycomb Hofstadter model, with nearest-neighbor interactions and $L_y=6$, we have $d=2$ and $D=23$, whereas for the full moir{\'e} superlattice Hamiltonian from Eq.~\ref{eq:many_ham} at the same cylinder circumference we have $d=4$, $D=105$, and twice as many sites in the MPS unit cell (due to the orbital degree of freedom). Therefore, at sufficiently large MPS bond dimensions, the simulations in this paper take approximately 36 times longer and require nine times more memory than comparable simulations of the Hofstadter model. In practice, these computations run slightly ($\lesssim10\%$) faster due to numerical optimizations in \textsc{LAPACK}, which is why these values are upper estimates.

As a result of the computational expense of these calculations, we are not able to reach MPS bond dimensions $10^3\lesssim\chi\lesssim10^4$ that are typical for modern iDMRG studies of Hofstadter~\cite{Schoonder19} or Hubbard~\cite{Szasz18} models. Hence, particular care is needed in analyzing the errors of our results. For each two-site iDMRG run, we set a desired entropy and energy error $\delta S = \delta E = 10^{-6}$, since we observe that the maximum error in the wave function and energy due to truncation of the two-site update is typically $|| \ket{\psi} - \ket{\psi_\text{trunc}} ||\sim 10^{-4}$ and $|| \braket{\psi|H|\psi} - \braket{\psi_\text{trunc}|H|\psi_\text{trunc}} ||\sim 10^{-4}$~meV, respectively. Here, we denote $\ket{\psi}$ as the effective wave function at the end of the Lanczos algorithm and $\ket{\psi_\text{trunc}}$ as the $\ket{\psi}$ after truncation to a given MPS bond dimension. By the end of all runs, the deviation of our MPS from canonical form is $\lesssim10^{-10}$, which verifies the translational invariance. 

\begin{figure}
	\includegraphics[width=\linewidth]{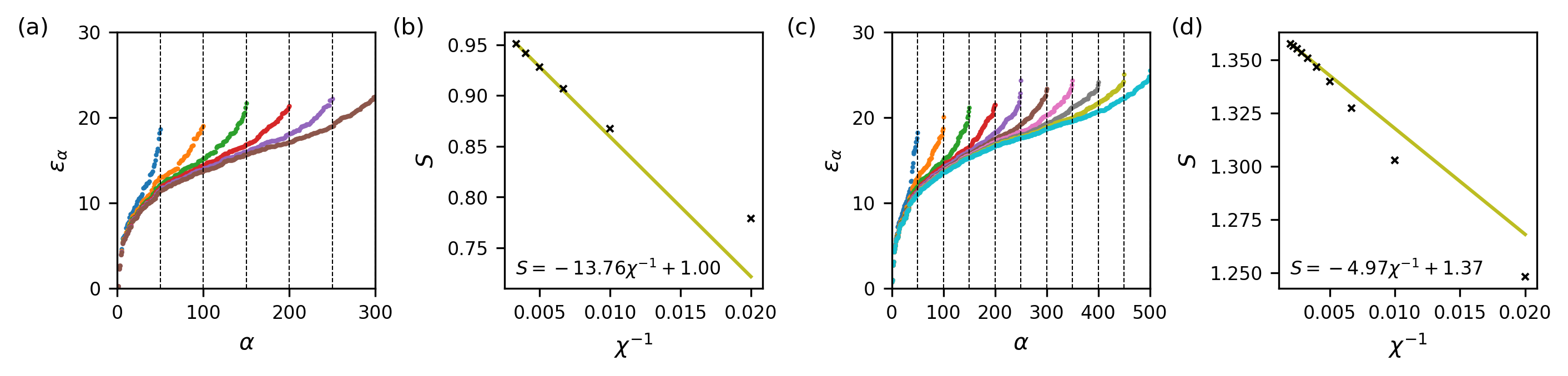}\\
	\caption{\label{fig:schmidt_values} Convergence analysis for [(a),(b)] the full moir{\'e} superlattice Hamiltonian given in Eq.~\ref{eq:many_ham} and [(c),(d)] the single-orbital fermionic honeycomb Hofstadter model. [(a),(c)] Entanglement energy $\epsilon_\alpha\equiv-\ln\Lambda_\alpha^2$ plotted against Schmidt value index, $\alpha$, for various bond dimensions $\chi$ indicated by their cut-offs. The system parameters are those for (a) the $n_\phi=1/3$, $L_y=6$ data point in the scaling plot shown in Fig.~\ref{fig:Ly_flow}(a) and (c) the $n_\phi=1/3$, $L_y=6$ Hofstadter model corresponding to a single-orbital copy of $H_0$ (Eq.~\ref{eq:single_ham}) at $\kappa=0$. [(b),(d)] Corresponding entanglement entropy $S=-\sum_\alpha \Lambda_\alpha^2 \ln \Lambda_\alpha^2$ against inverse bond dimension, with a line of best fit drawn through the last three points.}
\end{figure}

To quantify the convergence of our runs, we examine the convergence of the entropy from bipartitioning the cylinder in Fig.~\ref{fig:Ly_flow}, since the entropy is known to be among the most difficult quantities to converge in DMRG and a measurement of the topological entanglement entropy requires the highest accuracy from our results. In Fig.~\ref{fig:schmidt_values}(a) we plot the entanglement energies for various bond dimensions, and in Fig.~\ref{fig:schmidt_values}(b) we plot the convergence of the entanglement entropy. As we increase the bond dimension, the values of the entanglement energies converge, starting from the largest Schmidt value. Consequently, our estimate of the entanglement entropy ($S=-\sum_\alpha \Lambda_\alpha^2 \ln \Lambda_\alpha^2$) improves. From this, it is possible to estimate the error in the entanglement entropy by extrapolation. We note that our results in Fig.~\ref{fig:Ly_flow}(a) have converged sufficiently to verify the topological entanglement entropy to within $\delta\gamma=0.09$, and that other results, such as the flux insertion or the scaling with $\kappa$, do not require such accuracy to discern the topological features. For the single-orbital fermionic Hofstadter model with nearest-neighbor interactions, for example, we observe the correct charge pumping and spectral flow with $\chi=50$, even though it can be seen in Figs.~\ref{fig:schmidt_values}(c,d) that the entanglement entropy has not yet converged for this value of $\chi$.

\twocolumngrid
\bibliographystyle{apsrev4-1}
\bibliography{tbg_rev2}

\end{document}